\documentstyle[twoside,epsf]{article}
\input ibvs2.sty
\input ibvst.sty
\begin{document}
\IBVShead{6xxx}{xx Month 2014}

\IBVStitletl{ Resolved photometry of the binary components of RW Aur}

\IBVSauth{Antipin, S.; Belinski, A.; Cherepashchuk, A.; Cherjasov, D.;
Dodin, A.; Gorbunov, I.; Lamzin, S.; Kornilov, M.; Kornilov,V.; 
Potanin, S.; Safonov, B.; Senik, V.; Shatsky, N.; Voziakova, O.}

\IBVSinst{ Sternberg Astronomical Institute, Lomonosov Moscow State University, Russia. 
kolja@sai.msu.ru}

\begintext

\section{Introduction}

  RW Aur is one of objects from the initial list of T Tauri type stars
composed by Joy (1945).  Joy \& van Biesbroeck (1944) have discovered that
the star has a companion RW Aur B, which was at that moment 1.5$^m$ fainter
than the main star RW Aur A.  Current position of the companion RW Aur B is
the following: $\rho \simeq 1.45^{\prime\prime},$ PA$\simeq 256^\circ$
(Bisikalo et al., 2012).  It was found later that both A and B components
were classical T Tauri stars (Duch$\hat {\rm e}$ne et al., 1999), i.e. 
pre-main sequence low mass stars surrounded by accretion disks.  Spectral
types of the main component and the companion are K1-K4 (Petrov et al.,
2001) and K5 (Duch$\hat {\rm e}$ne et al., 1999), respectively.

   Variability of RW Aur was discovered more than a century ago by
L.P.Ceraski (Ceraski, 1906).  Historical lightcurve of the star (Beck \&
Simon, 2001; Grankin et al., 2007; Rodriguez et al., 2013) reflects the
total brightness of both components due to their proximity.  In UBVRI
bands the star demonstrates irregular variability, amplitude of which
increases from I to U band, that is typical for classical T Tauri stars.  In
particular, average brightness of RW Aur during 1985-2003 period in the
V-band was near 10.5$^m$ with an average amplitude of seasonal variations
$\simeq 1.4^m$ (Grankin et al., 2007).  

  It is commonly accepted to interpret the variability of RW Aur as that of
the brighter component~A.  We found the only paper of White \& Ghez (2001) 
where a quantitative information on the brightness of RW Aur B in the
UBVR$_c$I$_c$ bands is presented (from November 9, 1994 HST observations).

 Petrov \& Kozack (2007) concluded that the brightness and color of RW Aur~A
are governed by variations of the circumstellar extinction rather than of
the accretion.  It looks strange because the inclination of RW Aur~A
disk midplane to the line of sight lies between 30$^\circ$ and 45$^\circ$
(Cabrit et al., 2006).  Unexpected confirmation of Petrov \& Kozack
conclusion appeared in 2010 when a long and deep dimming of RW Aur
happened.  The dimming had a depth of 2 magnitudes, a duration of 180
days and presumably was due to occultation of RW Aur~A by a dust cloud
(Rodriguez et al., 2013).

   The V-magnitude of RW Aur during the dimming event fell down to $\simeq
13^m,$ that is close to brightness of RW Aur B, so it is not clear what was
the real amplitude of RW Aur~A dimming.  It was not possible to answer this
question due to the lack of resolved photometry of this double system.

  According to the AAVSO database (http://www.aavso.org), in a period from
April 2011 to the end of April 2014 RW Aur demonstrated its usual (pre-dimming)
behavior, e.g.  its V magnitude varied in an irregular way around average
value 10.5$^m.$ Then the star was not observed till October,\,23 when it
appeared that RW Aur dimmed again down to V$\simeq 12.6^m.$

\section{Observations and results}

  Multicolor imaging of RW Aur was performed on November 13/14, 2014 with a
newly installed 2.5~meter telescope ($F_{equiv}=20$ m) of the Caucasus
observatory of Lomonosov Moscow State University at the mount Shatzhatmaz
(http://lnfm1.sai.msu.ru/kgo/) in course of test precommissioning
observations aimed to check the image quality provided by the instrument. 
The telescope was equipped with a mosaic CCD camera manufactured by Niels
Bohr Institute based on two E2V CCD44-82 detectors (pixel size 15$\mu m$) 
and a set of standard Bessel UBVR$_c$I$_c$ filters from Asahi
Spectra Co.

 In course of observations the image quality with FWHM between 0.5 and
0.7~arcsec was routinely obtained confirming the excellent optics quality of
the instrument, delivered by the REOSC company of Safran group, France
(Poutriquet et al., 2012).  The binary was clearly resolved (see Fig.\,1): 
the wings of a brighter component image contribute $<7\,\%$ to the central
intensity of a fainter component in all bands. 
This time of year at the site is known to be characteristic of exceptionally
stable atmospheric turbulence conditions (Kornilov et al, 2014), so this
result was not unexpected.  The exposure time varied from 5~sec in the I$_c$
band to 300~sec in the U band, the middle date of measurements is JD
2456975.56.


\IBVSfig{4cm}{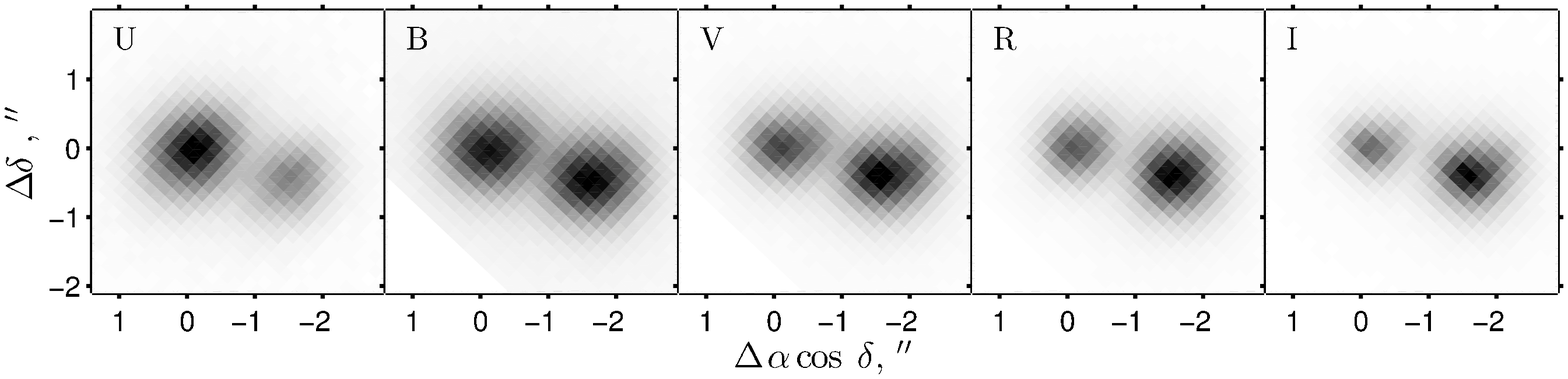}{Images of RW Aur binary in UBVRI
photometric bands.  The primary component RW Aur A is placed in the origin
of the coordinate system.}


  Primary data processing and PSF photometry were performed in a standard
way in the ESO-MIDAS environment with the DAOPHOT program package
(Stetson 1987).  Stars 127 and 129 from the AAVSO chart for RW Aur were used as
BVR$_c$I$_c$ photometric standards.  The U-B colors for these standards and
transformation of magnitudes and colors from the instrumental to the
standard UBVR$_c$I$_c$ system were made based on quasi-simultaneous
observations of Landolt standard fields (Landolt, 2009) using formulae from
Hardie (1964). The results of our measurements are presented in Table~1.

\begin{table}[h!]
\caption{UBVRI photometry of RW Aur}\label{ubvri-dat}
  \begin{tabular}{|c|c|c|c|c|c|}
\hline
 & U & B & V & R$_c$ & I$_c$ \\
\hline
RW Aur A & $14.26 \pm 0.3$ & $14.50 \pm 0.06$ & $ 13.80 \pm 0.05$ & $13.18 \pm 0.07$ & $12.46 \pm 0.1$
\\
\hline
RW Aur B & $14.97 \pm 0.3$ & $14.26 \pm 0.05$ & $ 12.92 \pm 0.03$ & $11.97 \pm 0.07$ & $11.01 \pm 0.1$
\\
\hline
  \end{tabular}
\end{table}


  The results are non-trivial, as follows from the comparison of our data
with that of White \& Ghez (2001) obtained 20 years ago (see Fig.\,2). 
First of all, during our observations RW Aur~A became $\simeq 3^m$ fainter
in all spectral bands (the dot-dashed curve at the left panel of the
figure) which may be interpreted as gray extinction.  A better fit can be
obtained assuming that the extinction is a sum of two components: a gray
extinction with A$_V=2.87$ and a selective standard one (Savage \& Mathis,
1979) with A$_V=0.44$ -- see open circles in the panel.  It seems natural to
explain current RW Aur A dimming as a result of eclipse of the star by dust
particles, with predominantly large enough size $r$ to produce gray
extinctions up to at least 0.7 $\mu m,$ which means that $r>1$ $\mu m$
(Kr\"ugel, 2003).


\IBVSfig{8cm}{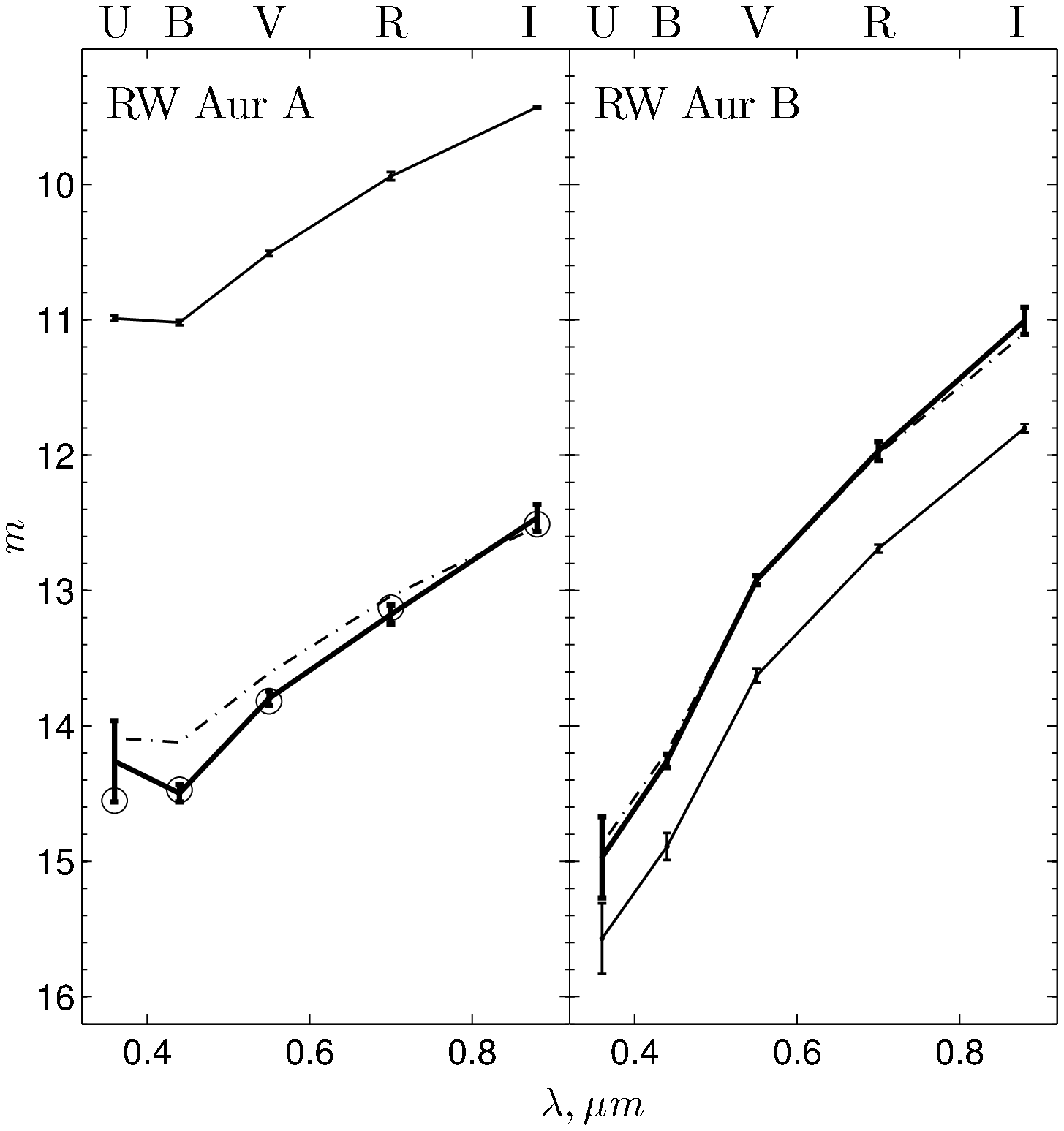}{UBVRI-photometry for A and B components of RW Aur
for two epochs: the thin lines are for HST observation (Nov.  1994), the
thick lines are for our observation.  The dash-dotted line corresponds to
the HST data shifted down by $3^m.1$ and up by $0^m.7$ for A and B
components, respectively.  The circles are obtained from HST data by applying
a sum of gray extinction with $\Delta m=2.87$ and selective extinction with
$A_V=0.44$ using a standard reddening curve.}


   Our results indicate that RW Aur B is also a variable star: at the moment
of our observations it was brighter than 20 years ago at $\Delta m \simeq
0.7^m$ in each of UBVRI band (gray brightening). Explanation is the same as
for RW Aur A, but in the opposite sense: in 1994, RW Aur B was eclipsed by
a cloud that consisted of dust particles with size $r>1$ $\mu m$ and now the
cloud has passed away from the line of sight.

   It follows also from our data that at the moment of observations the relative
contribution of RW Aur~B to the total brightness monotonically decreases from I
to U band: it dominates at long wavelengths but becomes fainter than RW Aur A
in the ultraviolet (see Fig.\,1).

  And last but not the least: our test observations indicate that the optics
of the new 2.5~m telescope is good as well as the seasonal astroclimate at the
site.

{\it Acknowledgements} We thank an anonymous referee for valuable
comments.  This research was carried out in the frame of Lomonosov Moscow
State University Program of Development.


\reference 

Beck T.L., Simon M., 2001, AJ, {\bf 122}, 413

Bisikalo D.V., Dodin A.V., Kaigorodov P.V. et al., 2012, Astron. Rep., {\bf
56}, 686

Cabrit S., Pety J., Pesenti N. and C. Dougados C., 2006, A\&A {\bf 452}, 897

Ceraski W., AN {\bf 170}, 339 (1906).

Duch$\hat {\rm e}$ne G., Monin J.-L., Bouvier J. and M$\acute {\rm e}$nard F., 1999,
A\&A, {\bf 351}, 954 (1999)

Grankin K., Melnikov S., Bouvier J. et al., 2007, A\&A {\bf 461}, 183

Hardie R.H., 1964, {\it Photoelectric Reductions} in the book {\it Astronomical
Techniques}, ed. W.A. Hiltner, University of Chicago, p. 178

Joy A.H., van Biesbroeck G., 1944, PASP, {\bf 56}, 123

Joy A.H., 1945, ApJ {\bf 102}, 168

Kornilov, V.,  Safonov, B.,  Kornilov, M., et al., 2014, PASP {\bf 126}, 482

Kr\"ugel E., {\it The Physics of Interstellar Dust}, 2003,
IoP Series in astronomy and astrophysics, Bristol, UK: The
Institute of Physics

Landolt A.U., 2009, AJ {\bf 137}, 4186

Petrov P.P., Gahm G.F., Gameiro J.F. et al., 2001, A\&A {\bf 369}, 993

Petrov P.P., Kozack B.S., 2007, Astron. Rep., {\bf 51}, 500

Poutriquet F., Plainchamp P., Billet J. et al., 2012, Proc. SPIE {\bf 8444}, 
Ground-based and Airborne Telescopes IV, 84441W

Rodriguez J.E., Pepper J., Stassun K.G. et al., 2013, AJ, {\bf 146}, 112

Savage B.D., Mathis J.S., 1979, Ann. Rev. A\&A {\bf 17}, 73

Stetson P.B., 1987, PASP {\bf 99}, 191

White R.J., Ghez A.M., 2001, ApJ, {\bf 556}, 265

\endreferences 

\end{document}